\documentclass[12pt]{iopart}
\usepackage{graphicx} 

\begin{document}
\title{Microwave single-tone optomechanics in the classical regime}

\author{Ilya Golokolenov$^{1}$, Dylan Cattiaux$^1$, Sumit Kumar$^1$, Mika Sillanp\"a\"a$^2$, Laure Mercier de L\'epinay$^2$, Andrew Fefferman$^1$ and Eddy Collin$^{1,*}$}
\address{$^1$ Univ. Grenoble Alpes, Institut N\'eel - CNRS UPR2940, 25 rue des Martyrs, BP 166, 38042 Grenoble Cedex 9, France}
\address{$^2$ Department of Applied Physics, Aalto University, FI-00076 Aalto, Finland}
\ead{eddy.collin@neel.cnrs.fr}
\vspace{10pt}
\begin{indented}
\item[]\today
\end{indented}
\begin{abstract}
We report on the quantitative experimental illustration of elementary optomechanics within the classical regime. 
All measurements are performed in a commercial dilution refrigerator on a mesoscopic drumhead aluminium resonator strongly coupled to a microwave cavity, using only strict single-tone schemes. 
Sideband asymmetry is reported using in-cavity microwave pumping, along with noise squashing and back-action effects.
Results presented in this paper are analysed within the simple {\it classical} electric circuit theory, emphasizing the analogous nature of classical features with respect to their usual quantum description.
The agreement with theory is obtained with {\it no} fitting parameters.
Besides, based on those results a simple method is proposed for the accurate measurement of the ratio between microwave internal losses and external coupling.
\end{abstract}

\submitto{\NJP}
\maketitle
\section{Introduction}
Cryogenic microwave technologies are widely used in applications dealing with superconducting quantum bits and quantum computing \cite{martinis, esteve}, which are today investigated in a continuously expanding number of laboratories. The development of 
scalable and robust quantum computing systems may find applications in very different fields, from numerical (quantum) simulations to cryptography.

Microwave nano-electro-mechanical systems (NEMS) are promising devices for such applications in quantum data processing \cite{silquan, cleland}. Such elements, which are based on the coupling of electro-magnetic waves with mechanical motion \cite{lehnert}, are proposed as new (quantum-limited) electric components \cite{JohannesFinkCirculator} or new (quantum-limited) sensors \cite{higginbotham}.
This principle was already foreseen in the 80s by C. Caves and V. Braginsky \cite{caves,braginsky}, and gravitational wave detection is indeed based on the fantastic sensitivity of opto-mechanics for motion detection \cite{ligo}.


Besides quantum electronics, microwave optomechanics could also lead to new components for {\it conventional} electronics, and experiments already demonstrated the feasibility of room temperature devices \cite{weig}.
For electronics engineers, the classical electric circuit modeling has been derived \cite{xin_arxiv}, which gives them the required tools for basic understanding and modeling of optomechanics. 

In the present article, we illustrate experimentally single-tone microwave optomechanics in the classical regime. The theoretical modeling of Ref. \cite{xin_arxiv} is applied with no free parameters. This demonstrates the validity of the classical approach, and also allows to apprehend classical phenomena analogous to quantum features discussed in the literature, like sideband asymmetry and back-action noise, without referring to the quantum operator formalism.
These effects will be presented with their classical understanding, and the link to quantum mechanics shall be addressed in the Discussion part of the manuscript.
We believe that our work will help to broaden the community of physicists developing the field of microwave optomechanics.



\section{Methods}

\begin{figure}[t]
\centering
\includegraphics[]{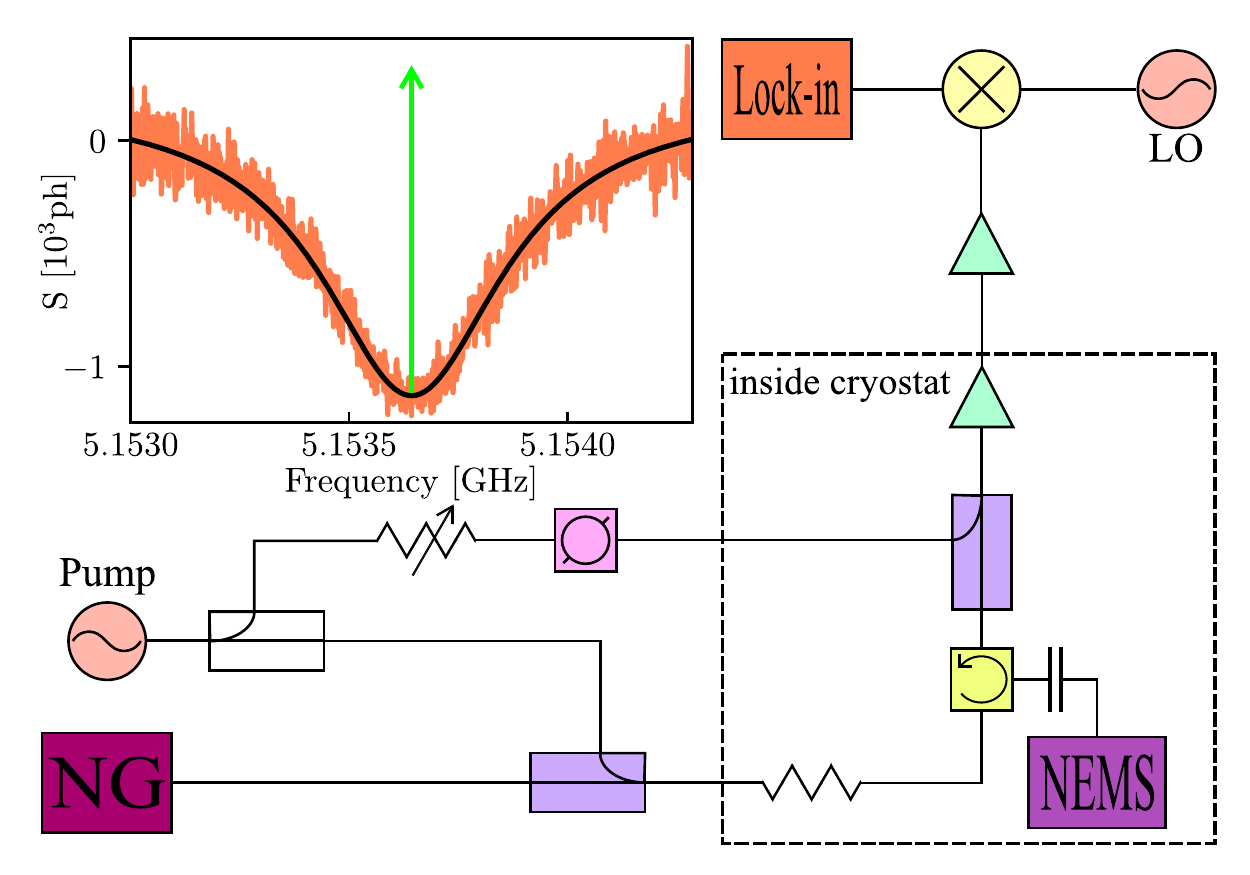}
\caption{{\bf Measuring scheme and cavity with applied noise.}
\newline
Signal from pump generator (Pump) goes to NEMS (embedded in on-chip microwave cavity) through two power combiners/dividers, a set of attenuators and a circulator, which leads to total losses of about $71 \ \mathrm{dB}$. 
A noise generator (NG) is used to apply external noise to the microwave cavity in a controlled way.
The NEMS sideband signal from the first mode is amplified by two HEMTs and measured by a Lock-in amplifier that demodulates at the frequency of a local oscillator (LO), with total gain about $48 \ \mathrm{dB}$. 
An amplitude-and-phase controlled cancellation line suppresses the pump before the first HEMT, in order to avoid its saturation at high powers. 
At the top left we show a cavity measurement $S(\omega)$ 
at full NG power applied (without pump tone; noise background removed for clarity). The green arrow shows the position of the pump in the case of in-cavity pumping, and the black line is a Lorentz fit with width $\kappa_{tot}/2 \pi$ (see text).
}
\label{fig1}
\end{figure}

A schematic of the setup is shown on fig. \ref{fig1}.
It enables standard single-tone microwave optomechanical measurements \cite{xin2019}, with the additional ability to artificially tune the microwave cavity effective temperature (or population in the quantum language) thanks to 
a Noise Generator (NG).
This NG is made of two High Electron Mobility Transistor amplifiers (HEMTs) in series (total gain $64 \ \mathrm{dB}$) connected on one side to a 50$~$Ohm load, and on the other to a $4-8 \ \mathrm{GHz}$ filter \cite{schwabnoise}. 
At the frequency of the cavity $\omega_c$, the applied microwave noise spectrum is essentially flat, with an amplitude $E_{noise}$ that can be quantitatively expressed in terms of photon number $E_{noise}/\hbar \omega_c$.
The noise level is tuned by inserting fixed attenuators, which lead to an uncertainty (from screwing/unscrewing) of about $\pm 1 \ \mathrm{dB}$ in noise level.

The basic measurement circuit is built around a circulator, two HEMTs in series and a lock-in detector (see fig. \ref{fig1}). A strong pump tone 
(of power $P_{in}$ on-chip) is sent together with the microwave noise
from NG
 into the cryostat,
and an amplitude-and-phase controlled cancellation line
allows to suppress it at the output, in order to avoid HEMT saturation. 
The sample is measured in reflection mode, and coupled capacitively to the line connected to the circulator. It consists of a microwave cavity in which a drumhead NEMS is embedded \cite{mikadrum}. 
The Brownian motion of the NEMS imprints sidebands in the cavity spectrum, which are measured with the lock-in that demodulates the signal with a Local Oscillator (LO) close to the pump frequency. 
With a broader span, the same technique enables to measure the cavity noise itself (see inset fig. \ref{fig1}); it can also be adapted to extract the linear response of the cavity, see Appendix.
In the present experiment, we study the first flexural out-of-plane mode of the drum.

There are 3 main schemes for single-tone optomechanics, namely: in-cavity (or nicknamed "green"), Stokes ("blue") and Anti-Stokes ("red") pumping \cite{AKM}. 

With in-cavity (or "green") pumping, the signal from the pump has a frequency equal to the resonance frequency of the cavity, $\omega_p = \omega_c$ (as depicted in fig. \ref{fig1} with green arrow).
The measured sidebands appear at frequencies $\omega_c \pm \omega_m$, where $\omega_m$ is the mechanical frequency of the drumhead resonator. 
The signal at $\omega_c - \omega_m$ will be further called "redside" signal, and at $\omega_c + \omega_m$ the "blueside" one. The two peaks are Lorentzian with a width given by the mechanical damping $\Gamma_m$. 

For Stokes pumping (or "blue amplification"), the pump signal frequency $\omega_p = \omega_c+\Delta$ is detuned from the cavity by $\Delta = + \omega_m$. 
The signal is then observed inside the cavity at $\omega_c$, the sideband at $\omega_c+2 \omega_m$ being strongly suppressed. 
This scheme is remarkable since it leads to {\it parametric amplification} of the Brownian motion: the linewidth of the peak narrows while its amplitude grows with applied pump power $P_{in}$.
This effect is called {\it dynamic backaction} \cite{AKM,xin_arxiv}.

With Anti-Stokes pumping (or "red cooling"), the pump signal is detuned from the cavity by $\Delta = - \omega_m$, and the sideband is also observed in the cavity (the other one being again suppressed). The remarkable feature of this scheme is that the dynamic backaction makes this time the damping increase with pump power  $P_{in}$, leading to {\it de-amplification} (or cooling) of the mode.


The aim of the present Article is to investigate single-tone optomechanics from a purely {\it classical electric} perspective \cite{xin_arxiv}.
Measurements were therefore done at $100 \ \mathrm{mK}$, which corresponds to a phonon population of about $140 \gg 1$, and the NG maximum noise applied at the sample level is about $1200  \gg 1$ photons, which is equivalent to about $300 \ \mathrm{K}$ (see inset fig. \ref{fig1}). 
In order to bridge classical and quantum mechanics, we shall express measured energy (power) in units of photons (photons/s) throughout the paper.

\section{Results} 

\subsection{In-cavity pumping}

To start with, the device was calibrated and characterised without applying external noise using the 3 standard schemes ("green", "blue" and "red" pumping).
From integrating the area of the measured sideband peaks appearing in the cavity spectrum $S(\omega)$ we obtain the {\it photon flux} $A$ (in photons/seconds) reaching the detector (see Appendix). From the classical electric circuit theory we have \cite{xin_arxiv}:

\begin{equation}
   A = \frac{k_B  T_{obs}}{\hbar \omega_m} g_0 ^2 n_{cav} \frac{\kappa_{ext}}{(|\Delta|-\omega_m)^2 + (\kappa_{tot}/2)^2} \, \frac{\Gamma_m}{\Gamma_m - \Gamma_{opt}\cdot sign(\Delta)} , \label{Seq}
\end{equation}

\begin{figure}[t]
\centering
\includegraphics[width=0.9\linewidth]{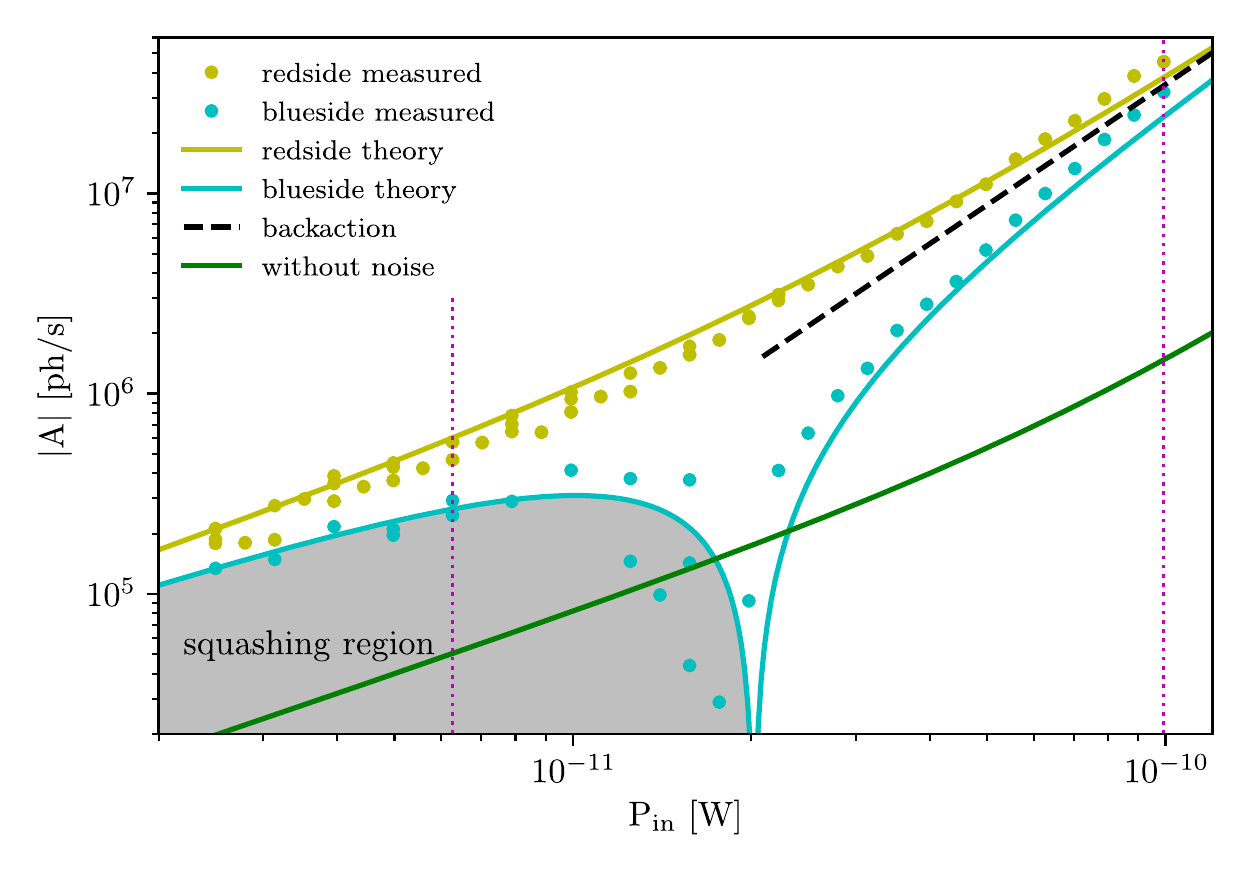}    
\caption{{\bf Power dependence of signal with in-cavity pump $\Delta =0$.} 
\newline
 Dots are measured signal (absolute value of sidebands area $|A|$) with 
 full noise from NG applied (about $1000$ photons),
 in cyan at $\omega_c+\omega_m$, the "blueside"; in yellow  at $\omega_c-\omega_m$, the "redside". Cyan and yellow solid lines are theoretical curves for "blueside" and "redside" signals respectively. 
 The dashed black line is a quadratic curve, showing the asymptotic backaction behavior. 
 The grey color filling marks the "squashing" region, where the signal is actually {\it negative}; the sharp feature in the cyan curve marks the changing of sign.
 The scatter there arises from the smallness of the signal. 
 The dashed magenta lines show powers at which noise cuts were made 
 (measurements at fixed $P_{in}$ vs. NG noise power, see fig. \ref{fig3} below).
 The solid green line corresponds to signals on both "blueside" and "redside" measured without external noise 
 (see text, and Appendix for corresponding data).}
\label{fig2}
\end{figure}

where: 
\begin{equation}
    n_{cav} = \frac{\kappa_{ext} P_{in}}{[\Delta^2 + (\kappa_{tot}/2)^2]\hbar \omega_c} ,
     \label{ncav}
\end{equation}

\begin{equation}
    \Gamma_{opt} = \frac{4g_0 ^2 n_{cav}}{\kappa_{tot}} ,
    \label{gammaopt}
\end{equation}
with $n_{cav}$ the cavity drive population, and $\Gamma_{opt}$ the optical damping. $\Delta = 0,+\omega_m,-\omega_m$ is the detuning of the pump from the cavity, for "green", "blue" and "red" schemes respectively.
The function $sign(\Delta)=-1,1$ and $0$ for negative, positive and zero $\Delta$ values respectively.
While the measurements are quoted in photons and photons/s (which is, in essence, what optomechanics is about), 
the classical nature of the theoretical approach shall lead us to discuss the results in terms of temperatures: 
$T_{obs} = T_{cryo} + T_{heat} $ is the observed temperature for the studied mode, with $T_{cryo}=100 \ \mathrm{mK}$ for our experiment. The $T_{heat} $ contribution comes from the backaction due to out-of-equilibrium photons populating the cavity because of the finite phase noise of the microwave generator used for the pump (see Appendix for details). 

In Eq. (\ref{Seq}), the last term comes from the dynamic backaction contribution which depends on the scheme through the $sign(\Delta)$ function \cite{xin_arxiv}. The middle term is the cavity susceptibility, with $\kappa_{tot}=\kappa_{ext}+\kappa_{in}$ its total damping rate and $\kappa_{in}$ the internal losses. $\kappa_{ext}$ is due to the capacitive coupling of the cavity to the measurement line, and $g_0$ (the {\it optomechanical coupling} in Rad/s) is obtained from the capacitive coupling between microwave and mechanical resonators. More details can be found in Ref. \cite{xin_arxiv} and Appendix. 

The fit parameters for this device are (see Appendix for details):

$g_0 = 2 \pi \cdot 220 \ \mathrm{Hz} \ (\pm 50 \ \mathrm{Hz})$ is the optomechanical coupling,

$\kappa_{ext} = 2 \pi \cdot 235 \ \mathrm{kHz} \ (\pm 5 \ \mathrm{kHz})$;
$\kappa_{tot} = 2 \pi \cdot 500 \ \mathrm{kHz} \ (\pm 50 \ \mathrm{kHz})$ are the cavity damping rates,

$\Gamma_m = 2 \pi \cdot 420 \ \mathrm{Hz} \ (\pm 50\ \mathrm{Hz})$ is the mechanical damping rate,

$\omega_m = 2 \pi \cdot 15.129 \ \mathrm{MHz} \ (\pm 100 \ \mathrm{Hz})$ is the mechanical frequency,

$\omega_c = 2 \pi \cdot 5.15365 \ \mathrm{GHz} \ (\pm 50 \ \mathrm{kHz})$ is the cavity frequency,

which define a quality factor of about $10^4$ for the cavity, and for the drumhead resonator of about $5\cdot 10^4$. 
Note that for this device $\kappa_{ext} \approx \kappa_{tot}/2$; we shall come back to this point in the following.

The theoretical curves corresponding to measured photon flux without added noise 
from NG
are shown as the full green curve in fig. \ref{fig2} (identical for the two sidebands, note the slight curvature at high powers arising from the pump noise), and the dashed blue and red lines in figure \ref{fig4} of the next Section (for the two other pumping schemes; 
see Appendix for the corresponding data). 

\begin{figure}[t]
\includegraphics[width=1\linewidth]{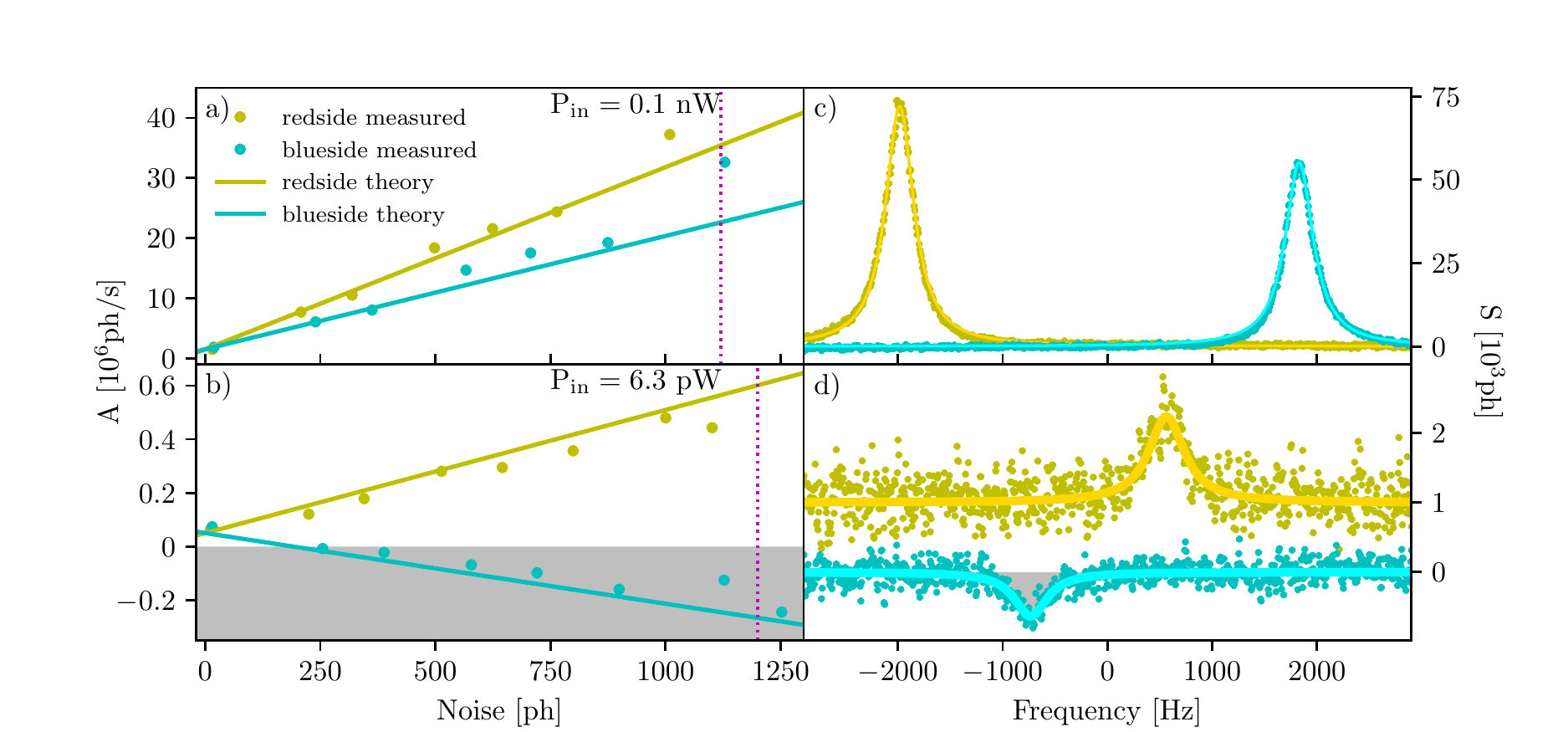}
\caption{{\bf Noise cuts and examples of measured signals} 
\newline
{\bf a)} and {\bf b)} show the photon flux $A$ dependence of both sidebands as a function of applied noise
from NG at two different powers.
For {\bf a)} {\it back-action} is dominating, and for {\bf b)} we observe the so-called {\it sideband asymmetry}. 
The grey filling color marks the noise squashing region, as in fig. \ref{fig1}, where the signal becomes negative.
{\bf c)} and {\bf d)} show examples of measured spectra $S$, where dots correspond to actual data and solid lines to Lorentzian fits.
Resonance frequencies in {\bf c} and {\bf d} appear slightly shifted because of the small residual "optical spring" effect due to imperfect pump tuning \cite{AKM}. 
Yellow points and curve in {\bf d} are shifted for clarity.
The magenta dashed lines define approximate noise values at which {\bf c)} and {\bf d)} were captured. 
}
\label{fig3}
\end{figure}

In fig. \ref{fig2} we plot 
the absolute value $|A|$ of
the pump power dependence of the two sidebands fluxes measured at full noise power from NG. In strong contrast with the initial (no noise, full green line) situation, now the "redside" (yellow) and "blueside" (cyan) sideband signals {\it do not} coincide anymore. This phenomenon is known as "sideband asymmetry", and has been already discussed in the framework of quantum mechanics \cite{asymmetry_first,asymmetry}.
Physically, it is due to the {\it interference} between the broadband microwave noise in the measuring (or output) port and the mechanically generated microwave noise in the sidebands; the sign of these correlations is reversed for these two (constructive or destructive interferences) \cite{xin_arxiv}.

At low powers, there is even a peculiar qualitative difference: for large enough noise levels from NG, the "blueside" peak in the spectrum even becomes {\it negative}: 
from a sideband peak, the signal evolves towards a sideband dip. This is the so-called phenomenon of "noise squashing" 
(when the destructive interference is even bigger than the Brownian noise level), already discussed in the literature within the quantum formalism \cite{squashing,schwabnature}. This is represented as a grey 
region in figs. \ref{fig2} and \ref{fig3}.


When microwave noise is added to the "green" scheme, 
the observed temperature $T_{obs}$ writes \cite{xin_arxiv}:
\begin{equation}
    T_{obs} = T_{cryo} + T_{heat}+ T_{sa} + T_{backaction}, \label{Tobs}
\end{equation}
with the contribution $T_{sa}$ depending on the measured sideband:
\begin{equation}
    T_{sa}=\frac{\omega_m}{\omega_c} T_{noise} \mathrm{ \ - \ for \ "redside"}, \label{redside}
\end{equation}
\begin{equation}
    T_{sa}=-\frac{\omega_m}{\omega_c} T_{noise} \mathrm{ \ - \ for \ "blueside"},\label{blueside}
\end{equation}
where:
\begin{equation}
    T_{noise} = \frac{E_{noise}}{k_B}. \label{noiseT}
\end{equation}
Eq. (\ref{noiseT}) expresses the injected noise energy in terms of a bath temperature (in Kelvin).
It essentially corresponds to the effective temperature of the measurement port as seen from the cavity.
The last term in the $T_{obs}$ expression Eq. (\ref{Tobs}) corresponds to the noise from the cavity fed back to the mechanical mode by backaction:
\begin{equation}
    T_{backaction}=T_{cav} \frac{\omega_m}{\omega_c} \frac{ \kappa_{tot} ^2}{2\omega_m ^2} \frac{\Gamma_{opt}}{\Gamma_m} ,
\label{backact}
\end{equation}
where: 
\begin{equation}
    T_{cav} = \frac{T_{noise}\, \kappa_{ext} + T_{cryo} \, \kappa_{int}}{\kappa_{tot}} .
\label{equil}
\end{equation}
The cavity noise generates a force noise that drives the mechanical element, contributing thus to the overall Brownian motion observed through Eq. (\ref{backact}).
Eq. (\ref{equil}) simply reflects the fact that the cavity effective temperature is defined by the two baths to which it is connected (the internal degrees of freedom and the microwave port). We remind that $\Gamma_{opt}$ is proportional to the pump power $P_{in}$, see Eqs. (\ref{ncav},\ref{gammaopt}).

These expressions are shown in fig. \ref{fig2} at fixed 
(maximal, about 1000 photons) noise power 
from NG
as a function of pump power (full lines); they reproduce very well the measured data, demonstrating "sideband asymmetry" and "noise squashing" at low pump powers. At high powers, they capture the backaction contribution arising from the cavity noise \cite{backaction}, converging towards the asymptote (quadratic law, dashed black line). 
This arises when $T_{backaction} \gg T_{cryo}+T_{heat}+T_{sa}$, which is by itself proportional to $P_{in}$.
In fig. \ref{fig3} (a,b left panels) we plot the dependence of the measured signal $A$ with respect to injected noise level
from NG, 
at fixed pump power (corresponding to the dashed magenta lines in fig. \ref{fig2}; 
so-called noise cuts).
 The lines again correspond to the theory, and match the data with no free parameters.

\begin{figure}[t]
\includegraphics[width=\linewidth]{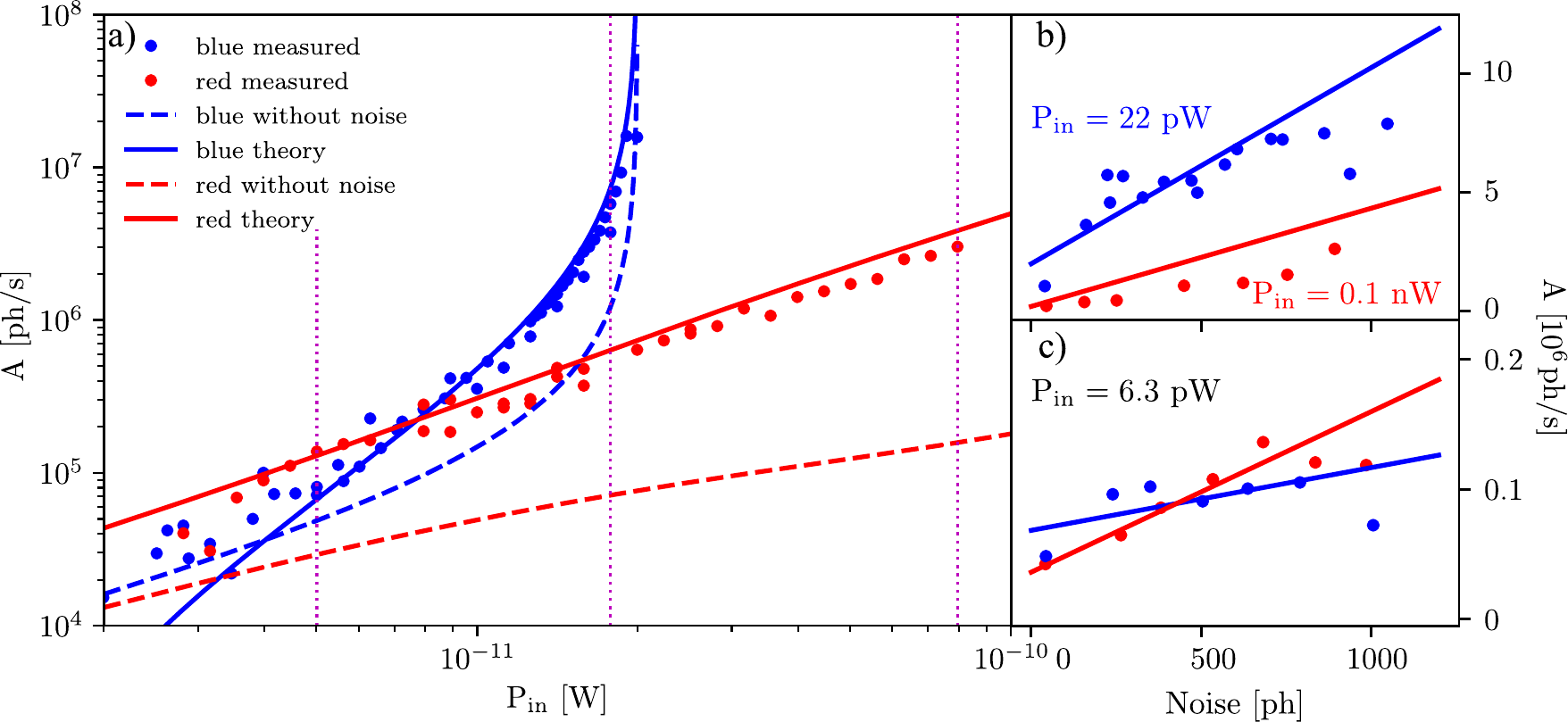} 
\caption{{\bf Power and noise cuts for "blue" and "red" pumping $\Delta =\pm \omega_m$} 
\newline
{\bf a)} shows the photon flux $A$ dependence on pump power with "blue" and "red" pumping schemes and maximum noise applied from NG.
{\bf b)} and {\bf c)} demonstrate the dependence on noise level at fixed pump powers (dashed magenta verticals in a). Dots are the data and solid lines theoretical predictions with no fitting parameters. 
Dashed red and blue lines mark the references without external noise (see Appendix). 
}
\label{fig4}
\end{figure}

On the right panels (c,d) of fig. \ref{fig3} we present actual measured resonance peaks, with their Lorentz fits, for two given pump powers and maximum noise from NG (dashed magenta lines at fig. \ref{fig3} a and b). "Asymmetry" and "squashing" are clearly identified in d, while the measured peaks are becoming very similar at the highest powers (c).
The asymmetry comes from the sign difference in $T_{sa}$ for "blueside" and "redside", Eqs. (\ref{redside},\ref{blueside}). In the extreme case where, on the "blueside", $|T_{sa}|> T_{cryo}+T_{heat}$ the peak becomes negative, which is the noise squashing effect.

\subsection{Stokes/Anti-Stokes pumping}

The same experiment can be conducted using "blue" or "red" sideband pumping schemes.
For "blue" and "red" the equations are quite similar, but with slightly different sideband asymmetry and backaction temperatures.
The theoretical expressions write \cite{xin_arxiv}:
\begin{equation}
    T_{sa}=\frac{\omega_m}{\omega_c} (2T_{cav} - T_{noise}) \mathrm{ \ - \ for \ "blue"},
\end{equation}
\begin{equation}
    T_{sa}=\frac{\omega_m}{\omega_c} (T_{noise} - 2T_{cav}) \mathrm{ \ - \ for \ "red"},
\end{equation}
\begin{equation}
    T_{backaction}=T_{cav} \frac{\omega_m}{\omega_c} \frac{\Gamma_{opt}}{\Gamma_m} .
\end{equation}
These expressions are plotted on experimental data in fig. \ref{fig4}, similarly to figs. \ref{fig2} and \ref{fig3}. The agreement is again very good, with no free parameters.

"Blue" and "red" schemes had been used in the past to demonstrate "sideband asymmetry" and "squashing" in microwave devices \cite{asymmetry,squashing,schwabnature}; but to our knowledge fig. \ref{fig2} is the first demonstration of these effects using the "green" pumping. 
Furthermore, our data is acquired within a genuine single-tone scheme: there is no "cooling tone" applied to bring the NEMS mode close to its ground state, since we want to keep it classical, in strong contrast with the previous references. 
Note the difference in the expressions of $T_{sa}$ for the different schemes.
One can see that for "blue" and "red" $T_{sa} \sim T_{noise}(2\frac{\kappa_{ext}}{\kappa_{tot}} -1 )$, which makes it very sensitive to $\frac{\kappa_{ext}}{\kappa_{tot}}$
(it even vanishes for $\frac{\kappa_{ext}}{\kappa_{tot}}=1/2$)
, while this is {\it not} the case for "green"; this shall be explicitly discussed in the next Section.

\section{Discussion}

The experiment is in good agreement with the classical electric circuit theory, with no free parameters: all terms are measured independently, as explained in the Appendix. This demonstrates the validity of the modeling, and validates its use as a new tool for engineers willing to develop {\it classical} applications of microwave optomechanics.

Besides, building on the formalism of Ref. \cite{xin_arxiv} our work illustrates 
classical analogs of "sideband asymmetry", "noise squashing", and "backaction force noise". These effects, which are in the literature discussed in terms of quantum operators, can be easily understood by electronics engineers with purely classical concepts.

Classical mechanics tells us that when the temperature of the measuring device, that is the microwave cavity $T_{cav}, T_{heat}$ and the microwave line $T_{noise}$, goes to zero, then $T_{obs}=T_{cryo}$: the NEMS temperature observed is exactly the cryostat temperature.
However, quantum mechanics tells us that this is impossible: an ultimate bound to back-action is imposed by the {\it zero point fluctuations} of the microwave mode. They will start to be relevant at low temperatures when the photon population in the cavity becomes close to 1; this is the natural limit of validity of the classical approach. 
Strictly speaking, the measured temperature $T_{obs}$ can never be exactly $T_{cryo}$, there will always be some addendum, no matter how small it might be. The precise definition of it constitutes the so-called {\it standard quantum limit}.
A closer look to the boundary between the classical electric circuit model and the quantum formalism can be found in Ref. \cite{xin_arxiv}.

Moreover, we can build on our results a method for measuring the $\frac{\kappa_{ext}}{\kappa_{tot}}$ ratio, {\it independently} of the usual microwave cavity calibrations. It is based on the high sensitivity of $T_{sa}$ for "blue" and "red" pumping to that ratio, while for "green" it is almost insensitive to it. 
Only when $T_{cav}=T_{noise}$ the $T_{sa}$ value is the same for the sideband signals measured in the different schemes;
the methodology requires thus to send a large microwave noise to the device, ensuring $T_{cav} \neq T_{noise}$. In this case, measuring the sidebands with the "green" scheme at small pump power one can calibrate $T_{noise}$ from the "sideband asymmetry". Measuring then the sidebands obtained using both "blue" and 'red" pumping, again at low powers, one can define $T_{cav}$, and therefore $\frac{\kappa_{ext}}{\kappa_{tot}}$. The back-action dominated range at large pump powers is then a self-consistency check ensuring that the fits are correct.
The method is fairly sensitive, since a $5~\%$ change in $\frac{\kappa_{ext}}{\kappa_{tot}}$ leads to a clearly visible change in our fits.
This scheme is rather straightforward, and essentially only limited by the maximum power that the NG can afford.

\section{Conclusion}

We presented a quantitative experimental illustration of elementary optomechanics within the classical regime. The results are analysed with the classical electric circuit model of Ref. \cite{xin_arxiv} {\it without} fitting parameters. 
All measurements were performed on a microwave-drumhead sample using only strict single-tone schemes, in a dilution cryostat at 100 mK (phonon population $\gg$ 1) with added microwave noise (photon population $\gg$ 1).

The key optomechanical features which are sideband asymmetry, noise squashing and back-action drive are reported, in particular using in-cavity microwave pumping. 
The quality of the agreement between theory and experiment demonstrates the validity of the approach, and makes it a very useful tool for electronics engineers. This work will prove to be relevant for the development of classical electronics applications of microwave optomechanics at room temperature.
Besides, based on the measurement of sideband asymmetry with the 3 single-tone schemes, a simple and precise method to quantify the ratio between microwave internal losses and the external coupling is proposed.

\section*{Acknoledgements}
I.G. and E.C. wish to thank B. Pigeau and J-P. Poizat for useful discussions.
We acknowledge support from the ERC CoG grant ULT-NEMS No. 647917 (E.C.), and StG grant UNIGLASS No. 714692 (A.F.). M.S. was supported by the Academy of Finland (contracts 308290, 307757, 312057), by the European Research Council (615755-CAVITYQPD), and by the Aalto Centre for Quantum Engineering. The work was performed as part of the Academy of Finland Centre of Excellence program (project 312057). We acknowledge funding from the European Union's Horizon 2020 research and innovation program under grant agreement No.~732894 (FETPRO HOT).
The research leading to these results has received funding from the European Union's Horizon 2020 Research and Innovation Programme, under grant agreement No. 824109, the European Microkelvin Platform (EMP).

\section*{Appendix}

\subsection{Circuit and cavity characterisation}

The losses (in the injection line) and the total gain (of the detection line) have been calibrated using a conventional Vector Network Analyser (VNA) prior to the experiment. They are known typically within $\pm 2 \ \mathrm{dB}$ ({\it absolute} error, obtained by comparing two cryogenic setups). The gain of the HEMT chain used for the noise generator (NG) has been measured in the same conditions.

\begin{figure}[h]
\includegraphics[width=\linewidth]{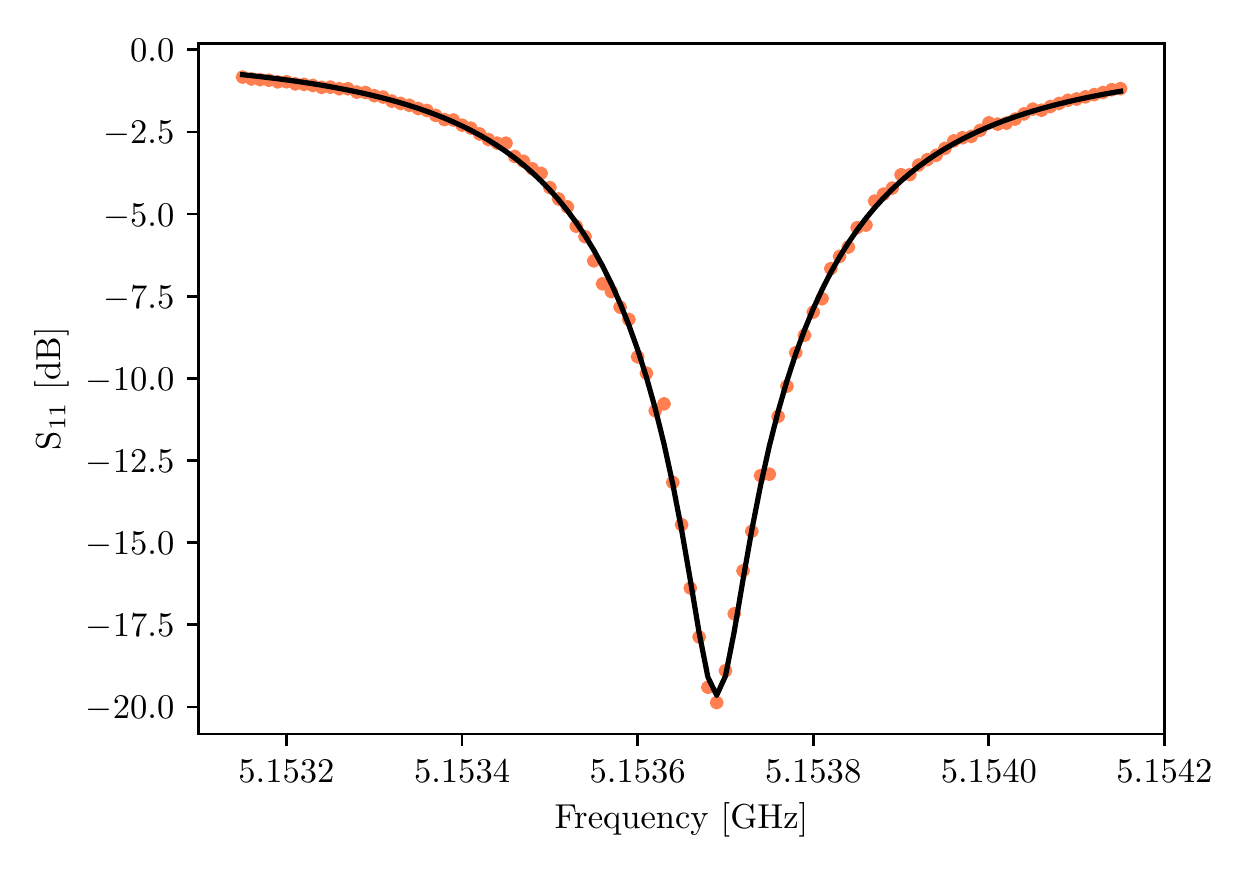}
\caption{{\bf Cavity measured with probe tone in reflection mode with its lorentzian fit (see text)}\newline 
Data taken at $100 \ \mathrm{mK}$, with a small probe power.
}
\label{cavitymes}
\end{figure}

The first step of the calibration procedure is then to characterize the microwave cavity. This is performed with a standard linear response procedure, applying a small probe tone (at frequency $\omega/ 2 \pi$, and fixed power $P_{probe}$) and measuring the output signal at the same frequency. Their ratio leads to the definition of the parameter $S_{11}[\omega]$, the reflection component of the {\it scattering matrix}. This can be fit by the expression \cite{AKMbook}:
\begin{equation}
    S_{11}[\omega]_{dB} = 20\,log|\frac{(Q_{ext} + i Q_{im} - 2Q)(Q_{ext} + i Q_{im})^{-1} + 2 i Q\frac{\omega - \omega_c}{\omega_c}}{1 + 2iQ\frac{\omega - \omega_c}{\omega_c}}| .
\end{equation}
In this expression, $Q_{im}$ captures the phase-dependent background that is hidden in a global amplitude measurement. Such a fit is shown in fig. \ref{cavitymes}.

We verified that the parameters are essentially independent of temperature below $250 \ \mathrm{mK}$, and independent of power $P_{probe}$ in the range of interest.
From the fit the following parameters are found: $\omega_c = 2 \pi \cdot 5.15365 \ \mathrm{GHz}$, $Q = 10^4$, $Q_{ext} = 22540$ which leads to $\kappa_{ext} = 2 \pi \cdot 230 \ \mathrm{kHz}$. The quoted error bars given in the main text are essentially due to the quality of the fit.

\subsection{3 schemes without noise}

The data from the lock-in is fitted with a Lorentzian expression (see at fig. \ref{fig3} (c, d) for the mechanical sidebands and inset in fig. \ref{fig1} for the cavity spectrum):
 \begin{equation}
     R[f] = \frac{a (\gamma/2)^2}{(\gamma/2)^2 + (f - f_0)^2} ,
 \end{equation}
 where $\gamma$  is the full width of the Lorentzian, $a$ its height and $f_0$ its position. 
 
\begin{figure}[h]
\includegraphics[width=\linewidth]{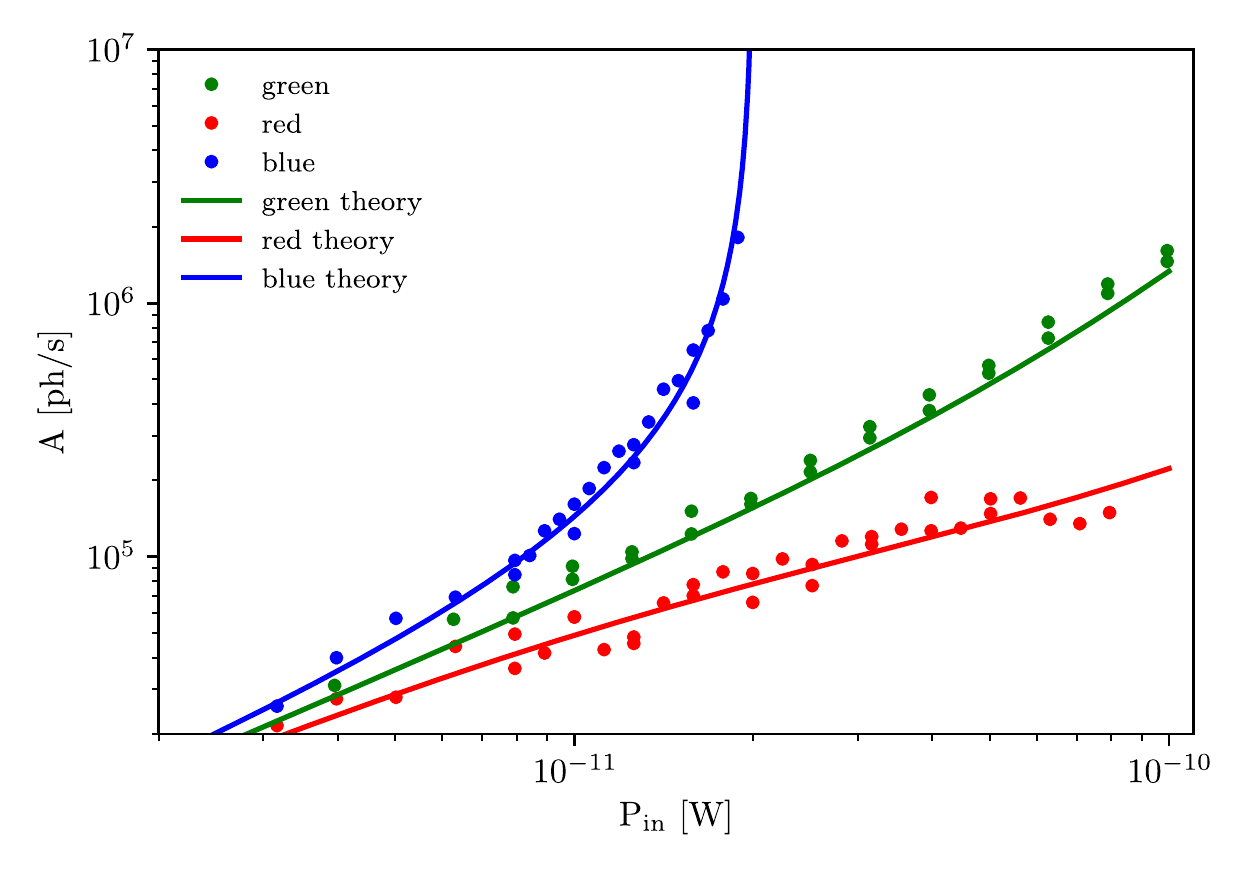}
\caption{{\bf Power dependence of signal $A$ with "green", "blue" and "red" pumping without added noise
} 
\newline
Standard measurements of signal area $A$ power dependence for the 3 single-tone schemes compared with theory [see Eq. (\ref{Seq})], at $100 \ \mathrm{mK}$. The signal with "green" pumping was acquired at both sidebands, the difference being barely visible and in the range of error bars.
}
\label{figap1}
\end{figure}

The Lorentzian fit of the cavity with full applied noise (fig. \ref{fig1}) leads to the number of noise photons $E_{noise}/\hbar \omega_c$, obtained from the height $a$.
For the sideband peaks, the fit gives the mechanical (effective) damping rate $\Gamma_m$ or $\Gamma_m \pm \Gamma_{opt}$ (depending on the scheme, see next Section below)  from $\gamma$, the mechanical frequency $\omega_m$ from $f_0$ and the peak area as:
\begin{equation}
     A = \pi a(\gamma/2)   \label{area} .
\end{equation}

The next step of the calibration procedure consists in verifying the 3 single-tone schemes {\it without} any added noise from NG. This is shown in fig. \ref{figap1}, plotting the area obtained from Eq. (\ref{area}) as a function of pump power.
The data is fit to Eq (\ref{Seq}), defining thus the last experimental parameter relevant to the experiment: $g_0 = 2 \pi \cdot 220 \ \mathrm{Hz}$. The error bar on this value (also obtained from the linewidth plot, see next Section) given in the main text is essentially due to the 
uncertainty in the losses/gains calibrations.
Besides, a small "out-of-equilibrium heating" contribution due to the phase noise of our generator is visible for the "green" pumping scheme (see deviation from linear behaviour in fig. \ref{figap1}). We fit it to be 
  $T_{heat} = 9\cdot 10^{11} P_{in} ^{1.3}$ (in K, for $P_{in}$ in W), in a similar fashion to Ref. \cite{xin2019}. 
For "blue" and "red" pumping schemes, an equivalent fit is used with a 
numerical prefactor of $5 \cdot 10^{11}$.
On the other hand, there is no {\it physical} heating, caused by the applied pump power, measurable in this range of parameters. This is not always true for this kind of devices, see e.g. Ref. \cite{prresearch}. 

\subsection{Damping power dependence with and without noise}
\begin{figure}[h]
\includegraphics[width=\linewidth]{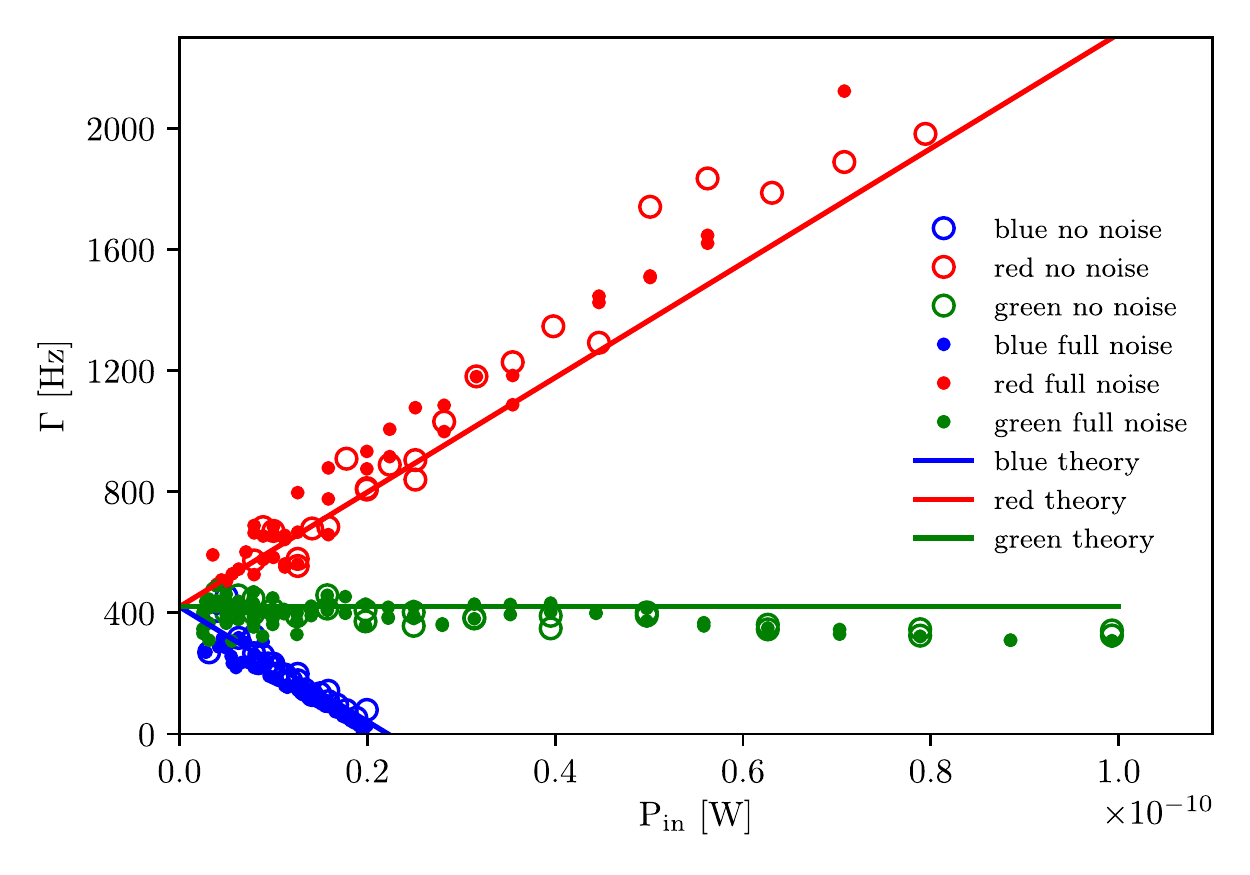}
\caption{{\bf Damping power dependence with and without noise} 
\newline 
Data taken at $100 \ \mathrm{mK}$ from the full linewidth of the peak. The measurements without noise correspond to the same batch as fig. \ref{figap1}. For the "green" pumping scheme, the two sidebands give the same value, within error bars (see scatter). 
}
\label{figap2}
\end{figure}

The mechanical damping $\Gamma$ is obtained from the linewidth of the measured peak. The data is shown in fig. \ref{figap2} for the 3 pumping schemes, as a function of pump power, {\it without} noise (same data sets as fig. \ref{figap1}) and {\it with full noise} applied. The theoretical fits read:
\begin{equation}
    \Gamma = \Gamma_m - sign(\Delta)  \Gamma_{opt} = \Gamma_m - sign(\Delta)  \frac{4g_0^2}{\kappa_{tot}} \frac{\kappa_{ext}P_{in}}{[\Delta^2 + (\kappa_{tot}/2)^2]\hbar \omega_c} , 
\end{equation}
which confirms the value of $g_0$ obtained in the previous Section.

 Fig. \ref{figap2} demonstrates that there is no difference between dampings in these two situations. We thus conclude that the added microwave noise does not modify the {\it dynamic part} of backaction.
 Besides, no signatures of actual heating of the sample related to the added noise could be measured, which proves that only optomechanics is involved.

\end{document}